\documentclass[11pt]{article}
\usepackage{comment}
\usepackage[margin=2.5cm]{geometry}
\usepackage{amsmath,amssymb,extarrows,mathtools,graphicx,subfigure,setspace,bbold}
\usepackage{cite}
\usepackage{slashed}
\usepackage[dvipsnames]{xcolor}
\usepackage{hyperref}
\hypersetup{colorlinks=true, linkcolor=blue, citecolor=magenta, linktoc=page}
\usepackage{tikz}
\usepackage{arydshln,leftidx,mathtools}
\usetikzlibrary{decorations.pathreplacing}
\numberwithin{equation}{section}
\newcommand{\be}{\begin{equation}}
\newcommand{\bea}{\begin{eqnarray}}
\newcommand{\eea}{\end{eqnarray}}
\newcommand{\ba}{\begin{align}}
\newcommand{\ea}{\end{align}}
\newcommand{\ee}{\end{equation}}
\newcommand{\nn}{\nonumber}

\begin{document}

\begin{titlepage}
\thispagestyle{empty}

\begin{flushright}
IPM/P-2017/010\\
\end{flushright}

\vspace{.4cm}
\begin{center}
\noindent{\Large \textbf{Holographic Subregion Complexity for Singular Surfaces}}\\
%\noindent{\Large \textbf{Holographic Aspects of Theories with Momentum Relaxation}}\\
\vspace{2cm}
Elaheh Bakhshaei${}^{a}$, Ali Mollabashi${}^{b}$, Ahmad Shirzad${}^{a,c}$
\vspace{1cm}

${}^a$ \textit{Department of Physics, Isfahan University of Technology}
\\
\textit{P.O.Box 84156-83111, Isfahan, Iran}

\vspace{5mm}
${}^b$ \textit{School of Physics,}
%, Institute for Research in Fundamental Sciences (IPM),
%\\
%P.O.Box 19395-5531,Tehran, Iran 
\\${}^c$ \textit{School of Particles and Accelerators,}
\\\textit{Institute for Research in Fundamental Sciences
(IPM), P.O.Box 19395-5531, Tehran, Iran}
%\vspace{1cm}
%Emails: {\tt m$_{-}$mohammadi, mollabashi, and farzad@ipm.ir}

\vskip 2em
\end{center}

\vspace{.5cm}
\begin{abstract}
Recently holographic prescriptions are proposed to compute quantum complexity of a given state in the boundary theory. A specific proposal known as `holographic subregion complexity' is supposed to calculate the complexity of a reduced density matrix corresponding to a static subregion.
We study different families of singular subregions in the dual field theory and find the divergence structure and universal terms of holographic subregion complexity for these singular surfaces. We find that there are new universal terms, logarithmic in the UV cut-off, due to the singularities of a family of surfaces including a kink in (2+1)-dimension and cones in even dimensional field theories. We also find examples of new divergent terms such as squared logarithm and negative powers times the logarithm of the UV cut-off parameter.
%We show that there are different family of cu
%Our results show new divergence terms in comparison with smooth subregions for some singular surfaces with flat locus.
%In case of curved locus surfaces no new universal term appears.
%and even divergent terms similar to rececnt results for the universal terms of the another proposal known as `complexity=action' on a Wheeler-DeWitt patch.

\end{abstract}
%\keywords{Complexity,\ Holographic,\ Entangelment Entropy }

\end{titlepage}

\newpage

\tableofcontents
\noindent
\hrulefill

\onehalfspacing

%%%%%%%%%%%%%%%%%%%%%%%%%%%%%%%%
%%%%%%%%%%%%%%%%%%%%%%%%%%%%%%%%
\section{Introduction}
Quantum entanglement has been widely studied in the context of holographic field theories after the pioneering Ryu-Takayanagi (RT) proposal \cite{Ryu:2006bv, Ryu:2006ef}. Quantum complexity is another notion in quantum information theory which has been recently included in the context of holographic field theories. Roughly speaking, quantum complexity of a state is the minimum number of information gates needed to prepare a state from a given reference state. There exist some efforts to develop a holographic dual for quantities related to this notion in the context of AdS/CFT correspondence \cite{Susskind:2014rva, Stanford:2014jda, Susskind:2014moa, Brown:2015bva, Brown:2015lvg, Ben-Ami:2016qex, Couch:2016exn, Chapman:2016hwi, Carmi:2016wjl}.

From a more geometrical point of view, it is well-established that the von Neumann entropy of a subregion in a given state corresponds to the area of a co-dimension two surface in the gravity solution dual of the state. People have also tried to find geometrical duals for other quantities in the context of information theory; such as Renyi entropies \cite{Hung:2011nu, Dong:2016fnf}, information metric (fidelity susceptibility) \cite{MIyaji:2015mia, Alishahiha:2015rta, Banerjee:2017qti}\footnote{See also \cite{Bak:2015jxd}.}, fisher information \cite{Lashkari:2015hha}, etc..  Some of these geometrical objects are still co-dimension two objects in the dual theory but some are not.    

There are two distinct proposals to compute complexity of a state in the dual gravity theory. The first one, which is sometimes called the `complexity=volume' proposal, states that the complexity of a given state at a given time in the boundary theory is given by the volume of an extremal co-dimension one surface in the bulk which meets the corresponding time slice. To be more concrete, one can state this proposal as
\be\label{CV}
\mathcal{C}_V=\mathrm{max}\left[\frac{V}{G_N\ell}\right],
\ee
where the maximum is chosen among those co-dimension one surfaces which end on the corresponding time slice on the conformal boundary. In this proposal $\ell$ is some length scale  which should be identified case by case, e.g.  the radius of the asymptotically AdS solution or the radius of the horizon in case of AdS black-hole geometries. This non recognized length scale seems to be a disadvantage of this proposal. 

The other proposal, which is sometimes called `complexity=action', states that the complexity of a given state at a given time is equal to the on-shell action of the dual (Einstein) gravity theory computed in the domain of dependence of any Cauchy surface in the bulk which ends on the given time slice at the conformal boundary.\footnote{Recently some progress have been made for complexity in higher derivative theories in \cite{Alishahiha:2017hwg}.} This region is known as the Wheeler-DeWitt patch corresponding to the given boundary time slice. Although this proposal (in contrast with the previous one) does not need any length scale by definition, it has its own challenges due to surface terms and corner contributions of the Wheeler-DeWitt patch (see \cite{Lehner:2016vdi, Carmi:2016wjl}). We will come back to this point in the next section.

A natural generalization of the `complexity=volume' proposal concerns with generic mixed states. A specific way of constructing a mixed state out of the entire state of a system is to trace out a part of the space-like manifold of the dual field theory. The mixed state constructed in this way is described by what is well-known as the reduced density matrix. Then the complexity of such a (static) state is proposed to be given by the volume enclosed by the Ryu-Takayanagi surface and the corresponding subregion in the boundary theory.\footnote{Recently a covariant generalization of this proposal is given in \cite{Carmi:2016wjl}.} To be more concrete the subregion complexity is defined as \cite{Alishahiha:2015rta}
\begin{align}
\mathcal{C}_{\mathrm{subregion}}=\mathrm{max}\left[\dfrac{V(\gamma)}{8\pi \ell G_N}\right], \label{b1}
\end{align}
where $\gamma$ is the RT surface of the corresponding subregion and $\ell$ is a length scale of the dual geometry. The maximization is among volumes enclosed by surfaces ending on the same subregion. This proposal (up to a numerical factor) reduces to `complexity=volume' given in \eqref{CV} if the subregion is chosen to be the whole time slice of the dual theory.

Different proposals for complexity all lead to UV divergent results since they all contain a volume of a surface which reaches the conformal boundary of an asymptotically AdS geometry. This is the same as what happened in the case of holographic entanglement entropy. Natural questions about such quantities are: ``What is the divergent structure of this quantity?", ``How it can be regularized?",  ``What kind of universal information can be extracted from it?", and ``Is it possible to find any monotonic function out of this quantity under the RG flow of the dual theory?". Specifically for the case of subregion complexity one may also ask about the (subregion) shape dependence of the divergence structure.

Some of the above questions has been recently addressed for different proposals of complexity and even for complexity of reduced states due to smooth subregions \cite{Carmi:2016wjl}. The goal of this paper is to investigate the divergence structure of subregion complexity when the subregion is a singular surface. Similar to the case of entanglement entropy we expect new divergent (sometimes new universal) terms due to singularities in the subregion. There has been done a considerable amount of efforts to investigate the role of singularities of entangling regions in the context of (mostly holographic) entanglement entropy \cite{Casini:2006hu, Casini:2008as, Hirata:2006jx, Myers:2012vs, Singh:2012ala, Bueno:2015rda, Bueno:2015xda, Pang:2015lka, Alishahiha:2015goa, Miao:2015dua, Bueno:2015qya, Bueno:2015lza, Myers:2012ed, Mozaffar:2015xue}.
We will consider the simplest case of a singular surface in a $(2+1)$-dimensional field theory and its generalizations to enough symmetric singular surfaces in higher dimensions (see \cite{Myers:2012vs, Singh:2012ala} for a similar analysis for entanglement entropy) and study the divergent structure due to subregion complexity proposal \cite{Alishahiha:2015rta}.

The rest of this paper is organized as follows: in section \ref{sec:2} we define different families of singular surfaces which we study. If the reader is just interested in the final results, we have summarized our subsequent results in this section. In the following sections we study complexity of different subregions and we finalize in the last section with addressing interesting  directions for future studies.

%%%%%%%%%%%%%%%%%%%%%%%%%%
%%%%%%%%%%%%%%%%%%%%%%%%%%
\section{ Singular Subregions and Summary of Results}\label{sec:2}
We are interested in asymptotically AdS solutions of Einstein gravity with a negative cosmological constant in $d+1$ dimensions. The simplest case which we study in this section is the pure $AdS_{d+1}$ solution in the Poincare patch with the following coordinates 
\begin{equation}
ds^2=\dfrac{L^2}{z^2}\left[-dt^2+dz^2+d\rho^2 + \rho^2 (d\theta ^2+ \sin ^2\theta d \Omega _n^2) +\sum _{i=1}^{m} (dx ^i)^{2}\right],\label{a1}
\end{equation}
where $z $ is the radial coordinate and $L$ is the AdS radius.
%\fixme{should be completed after the results in presence $f_1$ and $f_2$ are included.}
%which provides a natural length scale of the theory.
Here $ d\theta ^2+ \sin ^2\theta d \Omega _n^2$ is the metric on a unit sphere $S_{n+1}$ and the term $\sum _{i=1}^{m} (dx ^i)^{2}$ indicates a flat $R^m$ space in Cartesian coordinates. The conformal boundary of this solution is achieved in the $z \rightarrow 0$ limit. Hence, the boundary metric reads  
\begin{equation}
ds^2=-dt^2+d\rho ^2+ \rho^ 2(d\theta ^2+ \sin ^2\theta d\Omega _n ^2)+\sum _{i=1}^{m} (dx ^i)^{2}. \label{a2}
\end{equation}
For the whole manifold of the bulk, as well as the boundary, the range of the parameter $\theta$ is $(-\pi,\pi)$ for $n=0$ and $(0,\pi)$ for $n>0$. However, throughout this paper we consider different kinds of singular subregions, i.e.  the conic singular subregions, in which $-\Omega<\theta<\Omega$ for $n=0$ and $0<\theta<\Omega$ for $n>0$. The simplest conical geometry is a kink ($k$) in $d=3$ where $n=m=0$, 
	as the following subregion of the boundary
	\begin{align*}
	 k=\lbrace t_E=0,\rho=[0,\infty),-\Omega<\theta< \Omega\rbrace .
	\end{align*}
\begin{figure}\label{fig:kc}
\begin{center}
\begin{tikzpicture}[scale=1.7]
\draw [fill=blue!20!white] (0,0)--(3,0)--(4,1)--(1,1)--(0,0);
\draw [fill=white] (1.25,0)--(2,.7)--(2.05,0)--(1.25,0);
\draw [blue!60!black,<->,line width=.2mm](1.74,.45).. controls (1.85,.4) and (1.93,.4) .. (2.02,.45);
\draw [blue!40!black] (1.8,0.2) node {{\large $\mathbf{\Omega}$}};
%%%%%%%%%%%%%%%%%%%%%%%%%%%%%%%%%
\draw [fill=blue!10!white] (6+2,0)--(7.7+2,.5)--(7.7+2,2.7)--(6+2,2.5)--(6+2,0);
\draw [fill=blue!10!white] (6+2,0)--(7.5+2,-.5)--(7.5+2,2.5)--(6+2,2.5)--(6+2,0);
\draw [blue!80!white,densely dashed,-] (6+2,0)--(7.5+2,.44);
\draw [fill=blue!20!white] (6+2,0)--(7.5+2,.44)--(7.5+2,2.5)--(6+2,2.5)--(6+2,0);
\draw [blue!60!black,<->,line width=.3mm](7.5+2,-.5).. controls (7.8+2,-.2) and (7.85+2,.4) .. (7.7+2,.5);
\draw [blue!60!black,thick] (7.95+2,0) node {{\large $\mathbf{\Omega}$}};
%%%%%%%%%%%%%%%%%%%%%%%%%%%%%%%%%
\shade[top color=blue!30!white](5,1.8) arc (180:0:1cm and .25cm)--(6,-.2)--cycle;
\draw [](5,1.8) arc (180:360:1cm and 0.25cm) -- (6,-.2) -- cycle;
\draw [](5,1.8) arc (180:0:1cm and 0.25cm);
\draw [blue!60!black,thick] (6,.3) node {2{\large $\mathbf{\Omega}$}};
\draw [blue!60!black,<->,line width=.2mm](5.74++.12,.45-.35)..controls(5.85+.12,.5-.35) and (5.93+.12,.5-.35) .. (6.02+.12,.45-.35);
\end{tikzpicture}
\caption{Left: The blue plane represents a constant time slice of a $d=3$ CFT with a kink ($k$) entangling region on it. Middle: Conical entangling region in a $d=4$ CFT. Right: A crease ($k\times R^m$) entangling region as a direct generalization of the kink in higher dimensions.}
\end{center}
\end{figure}
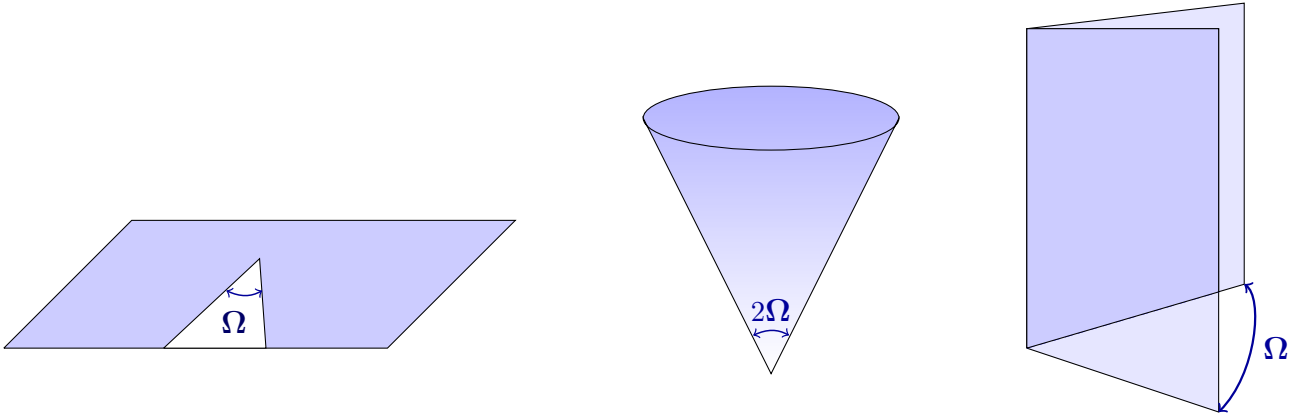
The cone family $(c_n)$ of singular surfaces in $d=n+3$ dimensions consists of manifolds with $m=0$ and $n\ge 1$ in Eq. (\ref{a2}) confined to the region 
\begin{align*}
c_n=\lbrace t_E=0,\rho=[0,\infty),\theta= \Omega \rbrace .
\end{align*}
The crease family in $d=3+m$ dimensions is the manifold $(k \times R^m)$ derived by considering $n=0$ and $m \ge 1$ in  Eq. (\ref{a1}).
We also consider mixed cases where both integers $n$ and $m$ are nonzero which we call them cone-crease.

We also study singular surfaces in asymptotically AdS$_{d+1}$ geometries given by
\begin{equation}
ds^2=\dfrac{L^2}{z^2}\left[dz^2+f_1(z)\left(-dt^2+d\rho^2 + \rho^2 (d\theta ^2+ \sin ^2\theta d \Omega _n^2)\right) +f_2(z)R^2 d \Omega _m^2\right].\label{a9}
\end{equation}
In these cases $f_1$ and $f_2$ are functions which are determined by the gravity equations of motion. We study different cones and creases in these asymptotically AdS geometries.

In Ref. \cite{Myers:2012vs} the holographic entanglement entropy for the above singular surfaces is calculated in Einstein gravity and also some specific higher derivative gravity theories. In this paper we calculate the holographic complexity in each case by using the proposal of Ref. \cite{Alishahiha:2015rta}. As we have mentioned in the previous section, according to this proposal the volume of a co-dimension one surface enclosed by the subregion in the boundary theory and the RT co-dimension two surface in the bulk is proportional to the complexity of the (mixed) state corresponding to the subregion.
%For this we  reason we need to calculate the bulk volume embraced by the Rio-Takayanagi surface for each of the problems.
To do so, one should find the RT surface corresponding to subregion $A$ which we denote by $\gamma_A$ and calculate the volume $V(\gamma_A)$ enclosed by $\gamma_A$. The holographic complexity is proposed to be given by Eq. \eqref{b1} \cite{Alishahiha:2015rta}. 
We choose $\ell$ in the asymptotically AdS gravity solutions to be identified with the AdS radius. In what follows we will study this quantity in different singular subregions.
	
\subsection*{Summary of Results}
Since the detailed calculations presented in next sections may be involved, here we briefly summarize our results. We study the divergent structure of holographic subregion complexity and find new divergences due to singular subregions which in some cases lead to new universal terms.

In the case of a crease entangling region in a (2+1)-dimensional boundary theory (see the left panel of Fig. \ref{fig:kc}) we find that there is a new divergent term of the form $\log\delta$ which is a universal term. The entanglement entropy for the same subregion also leads to a logarithmic universal term.

\begin{center}
\begin{tabular}{|c|c|c|c|c|c|}
\hline
$ d$ & Backgrround & Geometry of & Crease & Expected & New \\
 & spacetime & entangling surface & dimension  & divergences & divergences \\
\hline
3 & $R^3$ & $k$ & 0 & $1/\delta ^2$ & $\log \delta$\\
\hline
4 & $R^4$ & $c_1$ & 0 & $1/\delta ^3$ , $1/\delta$ & $\log \delta$\\
\hline
5 & $R^5$ & $c_2$ & 0 & $1/\delta ^4 , 1/\delta ^2 , \log \delta$ & $\log ^2 \delta$\\
\hline
6 & $R^6$ & $c_3$ & 0 & $1/\delta^5 , 1/\delta ^3 ,1/\delta$ & $\log\delta$\\
\hline
7 & $R^7$ & $c_4$ & 0 & $1/\delta ^6,1/\delta^4,1/\delta^2 , \log \delta$ & $\log ^2 \delta$\\
\hline
\hline
$>3$ & $R^d$ & $k\times R^{d-3}$ & $d-3$ &  $1/\delta^{d-1},  1/\delta^{d-3}$ & - \\
\hline
4 & $R^3\times S ^1$ & $k\times S^1$ & 1 &  $1/\delta^3$ & -\\
\hline
5 & $R^3\times S ^2$ & $k\times S^2$ & 2 &  $1/ \delta^4, 1/\delta^2,\log \delta$& -\\
\hline
6 & $R^3 \times S^3$ & $k\times S^3$ & 3 &  $1/\delta^5, 1/\delta^3$& - \\
\hline
6 & $R^4\times S^2$ & $k\times ( R^1\times S^2)$ & 3 & $1/\delta^5, 1/\delta^3,1/\delta$& -\\
\hline
\hline
5 & $R^5$ & $c_1\times R^1$ &1  & $1/\delta^4 , 1/\delta ^2 ,\log \delta$ & $1/\delta$\\
\hline
6 & $R^6$ & $c_1\times R^2$ &2 & $1/\delta^5 , 1/\delta ^3 ,1/\delta$ & $1/\delta ^2$\\
\hline
5 & $R^4\times S^1$ & $c_1\times S^1$ & 1 & $1/\delta^4, 1/\delta^2,\log \delta$ & $1/\delta $\\
\hline
6 & $R^4\times S^2$ & $c_1\times S^2$ & 2 & $1/\delta^5,1/\delta ^3,1/\delta$ & $1/\delta^2, \log\delta$\\
\hline
\hline
6 & $R^6$ & $c_2\times R^1$ &1 & $1/\delta^5 , 1/\delta ^3 ,1/\delta$ & $1/\delta \log\delta$\\
\hline
7 & $R^7$ & $c_2\times R^2$ &2 & $1/\delta^6 , 1/\delta ^4 ,1/\delta ^2, \log\delta $ & $1/\delta ^2\log\delta$\\
\hline
6 & $R^5\times S^1$ & $c_2\times S^1$ &1 & $1/\delta^5 , 1/\delta ^3 ,1/\delta$ & $1/\delta\log\delta$\\
\hline
\end{tabular}
\end{center}
\vspace{5mm}
For the case of a crease entangling region with a flat locus, which we denote by $k\times R^m$ (see the right panel of Fig. \ref{fig:kc}) there is no universal term due to the singularity and even no actual new divergent term, although the subleading divergent term gets corrections from the singularity. This resembles to the entanglement entropy in having no new universal term. Even for the case of $k\times S^1$, which again the locus of the singularity is flat, there is no new universal term and no new divergent contribution from the singularity.

In the case of creases with a curved locus we again find that there is no new divergent term. This is in contrast with what happens for entanglement entropy of these surfaces. We study the case of $k\times S^2$ and $k\times S^3$ and also  $k\times R\times S^2$  and in all of them although there is a $\log\delta$ term but it is suppressed with a positive power of $\delta$ resulting in no new divergent term.

The most interesting behavior happens for conical subregions which we show by $c_n$ (see the middle panel of Fig. \ref{fig:kc}). For these subregions we find that there is new universal $\log\delta$ term for odd $n$ and $\log^2\delta$ for even $n$'s. We have worked out a few examples of this for $n=1,2,3,4$. In comparison with entanglement entropy of these surfaces we find a shift from odd to even $n$'s where $\log^2\delta$ and $\log\delta$ appear respectively. It would be very interesting to find out whether these universal terms are related to some characteristic feature of the dual field theory. 

The other family of singular surfaces which we have studied are conical creases of the form $c_n\times R^m$ and $c_n\times S^m$. Among these surfaces the only case which we find that a universal $\log\delta$ term appears is $c_1\times S^2$. In other cases new divergent terms appear due to the singularity which have the form of $1/\delta\log\delta$ or $1/\delta^2\log\delta$. These are very similar to what has been recently found from the `complexity=action' proposal \cite{Carmi:2016wjl}. This similarity may be due to the singularities within the Wheeler-DeWitt patch. We have summarized our results in the above table.

%%%%%%%%%%%%%%%%%%%%%%%%%%%%%%%%%
%%%%%%%%%%%%%%%%%%%%%%%%%%%%%%%%%
\section{Flat Locus Singular Surfaces}
\subsection{Kink $k$}
The simplest case is a kink in a 2+1 dimensional boundary theory. The bulk metric dual to the vacuum state is given by
	\begin{equation}
	ds^2=\dfrac{L^2}{z^2}\left(-dt^2+dz^2+d\rho^2 + \rho^2 d\theta ^2\right),\label{a3}
	\end{equation}
and the subregion in defined in constant time slice as $ \rho \in  [0,H] $  and $ \theta  \in [-\Omega,\Omega]$, where $H$ is an IR cut-off. The corresponding Ryu-Takayanagi surface can be described by  $ z=z ( \rho,\theta )$, hence the entanglement entropy is given by
\begin{align}
S_{3}^{\mathrm{kink}}=\dfrac{2 \pi L^2}{l_{p}^2} \int d\rho d\theta\dfrac{ \sqrt{\rho ^2+\rho ^2 z'^2+\dot{z} ^2 }}{z^2}, \label{a4}
\end{align}
where $ z'=\partial_ {\rho} z$ and $ \dot{z}=\partial_ {\theta} z$. Since there is no length scale except $\rho$, the radial coordinate $z$ depends on $\rho$ linearly \cite{Hirata:2006jx}, i.e.  
\begin{align}
z=\rho \, h(\theta), \label{a5}
\end{align}
and $h(\theta)$ should be found such that it minimizes the entropy (area) functional and is anchored to the kink in the asymptotic boundary. Applying this into Eq. (\ref{a4}) gives
\begin{align}
S_{3,k}=\dfrac{4\pi L^2}{l_{p}^2} \int _{\delta / h_0}^{H}  \dfrac{d\rho}{\rho }\int_{0}^{\Omega - \epsilon} d \theta \dfrac{\sqrt{1+h^2+\dot{h}^2}}{h^2}, \label{a6}
\end{align}
where  $ \dot{h}=dh/d\theta $, $ h(0)=h_0 $ and $z=\delta $ is UV cut-off.
%Because of the symmetry under $\theta \rightarrow -\theta$  the maximum value of $h(\theta)$  occures at $\theta=0$ where $\dot{h}(0)=0$. In order to extrimize the surface, we should consider the Uler-Lagrange equation resulted from the functional  (\ref{a4}) for $ z(\rho ,\theta )$ and then substitute the ansatz (\ref{a5}) in it.
However, since the integrand of Eq. (\ref{a6}) does not depend on $\theta$ explicitly, we have the following conserved quantity along $\theta$ translation
\begin{align}
K=\dfrac{(1+h^2)}{h^2 \sqrt{1+h^2+\dot{h}^2}}=\dfrac { \sqrt{1+h_{0}^2}}{h_{0}^2}. \label{a7}
\end{align}
To find the holographic subregion complexity we should write the volume $ V(\gamma) $ of the subregion of the bulk
%embraced by the extrimal hyper-surface of Eq.  (\ref{a6}). Using the original metric (\ref{a3}) for the kink, the volume element of the bulk is $\dfrac{ L^3}{z^3}  d\rho \rho d\theta dz $. Hence, the corresponding volume can be written as
\begin{align}
\begin{split}
V(\gamma)&= L^3 \int_{\delta / h_0}^{H}d\rho \rho \int_{-\Omega +\varepsilon}^{\Omega -\varepsilon}d\theta \int_{\delta}^z \dfrac{dz}{z^3}\\
&=\dfrac{L^3}{2} \int_{\delta / h_0}^{H}d\rho \rho \int_{-\Omega +\varepsilon}^{\Omega -\varepsilon}d\theta \left(\dfrac{1}{\delta^2}-\dfrac{1}{ z^2}\right)\\
& = \dfrac{\Omega L^3}{2 \delta ^2}\left(H^2-\dfrac{\delta^2}{h_0^2}\right) -L^3\int_{\delta / h_0}^{H } \dfrac{d\rho}{\rho} \int _ 0 ^{ \Omega -\varepsilon}  \dfrac{d \theta}{h^2}, \label{a8}
\end{split}
\end{align}
where $\epsilon$ is a short distance cut-off in the boundary corresponding to $\delta$ in the bulk. To clarify the singular terms of Eq. (\ref{a8}) we  convert $\theta$ integration to an integral over $h$ as follows
\begin{align}
 \int _ 0 ^{ \Omega-\varepsilon} \dfrac{ d\theta}{h^2}=\int_{h_0}^{\delta  / \rho}\dfrac{d h}{h^2  \dot{h}}. \label{a9}
\end{align}
%From Eq. (\ref{a7}) and  noting that near the boundary $h$ is a  decreasing fuction (and hence $\dot{h}$  should be negative), we have 
One can easily find the following expression from Eq. (\ref{a7})
\begin{align}
\dot{h}=-\sqrt{\dfrac{(1+h^2)^2 h_0^4-h^4(1+h_0^2)(1+h^2)}{h^4(1+h_0^2)}}.\label{a10}
\end{align}
Using the coordinate transformation $ y=\sqrt{\dfrac{1}{h^2}-\dfrac{1}{h_0^2}}$, where $ y\rightarrow \infty $  as we approach the boundary via $\theta\rightarrow\Omega $, we have
\begin{align}
&\int_{h_0}^{\delta / \rho} \dfrac{d h}{h^2 \dot{h}}=\int _ 0^{\sqrt{(\rho  /\delta)^2-1/{h_0^2}}} d y\sqrt{\dfrac{(1+h_0^2)}{(1+h_0^2+y^2 h_0^2)(2+h_0^2+y^2 h_0^2)}}. \label{a11}
\end{align}
In the limit $ \delta\rightarrow 0 $ and hence $ y\rightarrow\infty $ the integrand is finite. So we can find it just for $ y\rightarrow\infty $. We have finally
\begin{align}
V(\gamma)=L^3\left[\frac{\Omega}{2}\dfrac{H^2}{\delta^2}+\alpha(h_0)\log\left( \dfrac{\delta}{H }\right)\right]+ \mathrm{finite}\ , \label{a12} 
\end{align}
 where $ \alpha (h_0) $ is the cut-off independent term given by
\begin{align}
\alpha(h_0)=\int _ 0^{\infty} d y\sqrt{\dfrac{(1+h_0^2)}{(1+h_0^2+y^2 h_0^2)(2+h_0^2+y^2 h_0^2)}}\ , \label{a13}
\end{align}
which vanishes in the smooth region limit (i.e. $\Omega\to\pi$).
Thus the divergent structure of holographic complexity of kink is given by
\begin{align}
\mathcal{C}_k=\frac{L^2}{8\pi G_N}\left[\frac{\Omega}{2}\dfrac{H^2}{\delta^2}+\alpha(h_0)\log\left( \dfrac{\delta}{H }\right)\right]. \label{a12} 
\end{align}

%%%%%%%%%%%%%%%%%%%%%%%%%%%%%
%%%%%%%%%%%%%%%%%%%%%%%%%%%%%
\subsection{Cone $c_n$}
As indicated in the previous section, to consider a conical subregion $c_n$ with $n=d-3$, we use the following form of the bulk metric
\begin{align}
ds^2=\dfrac{L^2}{z^2}\left[-dt^2+dz^2+d\rho ^2+ \rho^ 2(d\theta ^2+ \sin ^2\theta d\Omega _n ^2)\right], \label{a15}
\end{align}
where $d\Omega_n$ is the metric of a unit sphere $S_n$. The subregion in the boundary is defined by $ \rho \in  [0,H] $  and $ \theta  \in [0,\Omega]$. The extension of this region in the bulk is denoted by the function $ z(\rho ,\theta )$. One should find the profile of this extension via minimizing the following area functional
\begin{equation}
S= L^{d-1} \Omega_n  \int d\rho  d\theta\,\dfrac{\rho^{d-3}}{z^{d-1}}\,\sin^{d-3}\theta\,\sqrt{\rho ^2+\rho ^2 z'^2+\dot{z}^2}, \label{a16}
\end{equation}
where $ \Omega_n$ is the volume of the unit $n$-sphere and $ \dot{z}=\partial_ {\theta} z , z'=\partial _{\rho} z $.

%The equation of motion for $ z(\rho ,\theta )$ read
%******************************************************label   {a17}

As in the previous case, $z$ can depend on $\rho$ only linearly, i.e. $z(\rho,\theta) =\rho\, h(\theta)$. Using this assumption, and change of variable $ y=\sin\theta=y(h) $ which gives
$$ \dot{h} =\dfrac{\sqrt{1-y^2}}{y'}\;\;\;\;,\;\;\;\;\ddot{h}=-\dfrac{y {y'}^2 +(1-y^2) y''}{{y'}^3},$$
the equation of motion for the case $d=4$ read as follows
\begin{align}
0=& h (1+h^2) y (1-y^2)y'' -y y'\left( 3+h^2 +(3+5 h^2 +2 h^4){y'}^2\right)\nonumber\\
& +2h y^2\left(1+(1+h^2){y'}^2\right) -h\left(1+(1+h^2) {y'}^2\right)+(3+h^2)y^3 y' -h y^4,\label{a18}
\end{align}
where $ y'=\frac{dy}{dh} $ and $ y''=\frac{d^2 y}{dh^2} $. Since we are interested in the singular behavior of the complexity near the boundary, where  $ h\rightarrow 0 $, let us concentrate on this limit (still for $d=4$). For this reason we consider a power law expansion for $y(h)$ in terms of $h$ and put it in Eq. (\ref{a18}). Then using the boundary condition $y(0)= \sin \Omega$ we find the following result 
\begin{equation}
y=\sin (\Omega) -\dfrac{1}{4}\cos (\Omega) \cot(\Omega) h^2+\mathcal{O}\left(h^4\right).\label{a19} 
\end{equation}
The expansion for $\dot{h} $ follows consequently from  $ \dot{h} =\dfrac{\sqrt{1-y^2}}{y'(h)}$ as 
\begin{equation}
\dot{h}=-\dfrac{2\tan(\Omega)}{h} - \dfrac{1}{2}h(3-\cos(2\Omega)) \csc(2\Omega)\log(h)+\mathcal{O}\left(h\right).\label{a20}
\end{equation}
The corresponding volume is given by
\begin{align}
V(\gamma)&=L^4 \Omega _{1}\int d\rho \rho^{2}\int d\theta \sin (\theta) \int_{\delta} ^z \dfrac{dz}{z^4}\nn\\
&=\dfrac{L^4}{3}\Omega _{1}\int d\rho \rho^{2}\int d\theta \sin (\theta) \left(\dfrac{1}{\delta^{3}}-\dfrac{1}{z^{3}}\right)\nn\\
&=\dfrac{2\pi L^4 H^3}{9\delta ^3}(1-\cos(\Omega)) -\dfrac{L^4 2\pi}{3}\int _{\delta / h_0}^H\dfrac{d\rho}{\rho} \int_ {h_0}^{\delta / \rho} dh\dfrac{\sin(\theta)}{h^3\dot{h}}+\textnormal{ finite }.\label{a21}
\end{align}
Using asymptotic expansions (\ref{a19}) and (\ref{a20}) the integrand of (\ref{a21}) has the following behavior near the boundary
\begin{equation}
\dfrac{\sin(\theta)}{h^3\dot{h}}\sim -\dfrac{1}{2} \dfrac{\cos( \Omega)}{h^2}+\frac{1}{8}\cot ^2 (\Omega ) \sin (\Omega ) \csc (2\Omega)(3 - \cos (2\Omega)) \log (h)+\dfrac{1}{8}\cos(\Omega)\cot ^2(\Omega)+ \mathcal{O}\left(h\right).\label{a22}
\end{equation}
Let us divide singular parts of $V(\gamma)$ into $I_1$ and $L_2$ where the latter contains the singularities due to the integrand while the former shows the contribution of the limits of the integrations, i.e.
\begin{align}
I_1&= -\dfrac{L^4 2\pi}{3}\int _{\delta / h_0}^H\dfrac{d\rho}{\rho} \int_ {h_0}^{\delta / \rho} dh\Bigg[\dfrac{\sin(\theta)}{h^3\dot{h}} + \dfrac{1}{2} \dfrac{\cos( \Omega)}{h^2}  \nonumber \\
&\hspace{5mm} - \frac{1}{8}\cot ^2 (\Omega ) \sin (\Omega ) \csc (2\Omega)(3  - \cos (2\Omega)) \log (h)-\dfrac{1}{8}\cos(\Omega)\cot ^2(\Omega)\Bigg], \label{a23}\\
I_2&=\dfrac{2\pi L^4 H^3}{9\delta ^3}(1-\cos(\Omega))-\dfrac{L^4 2\pi}{3}\int _{\delta / h_0}^H\dfrac{d\rho}{\rho} \int_ {h_0}^{\delta / \rho} dh \Bigg[-\dfrac{1}{2}\dfrac{\cos( \Omega)}{h^2} \nonumber \\
&\hspace{5mm}+\frac{1}{8}\cot ^2 (\Omega ) \sin (\Omega ) \csc (2\Omega)(3 - \cos (2\Omega)) \log (h) 
+\dfrac{1}{8}\cos(\Omega)\cot ^2(\Omega)\Bigg].\label{a24}
\end{align}
 So the singular part of the complexity is given by
\begin{equation}
\mathcal{C}_{4,c_1}=\dfrac{1}{8\pi L G_N}( I_1+I_2).\label{a25}
\end{equation}
In the limit $h\rightarrow \delta/\rho$ there is no singular term from integration over $h$ (neither from the integrand nor from the integration limits); we have just a logarithmic singularity from the lower limit of the integration over $\rho$ as follows 
\begin{align} 
I_1=&\dfrac{L^4 2\pi}{3} \log \delta  \int_ {h_0}^ 0 dh\left(\dfrac{\sin(\theta)}{h^3\dot{h}} +\dfrac{1}{2} \dfrac{\cos( \Omega)}{h^2}-\frac{1}{8}\cot ^2 (\Omega ) \sin (\Omega ) \csc (2\Omega)(3 - \cos (2\Omega)) \log (h)-\dfrac{1}{8}\cos(\Omega)\cot ^2(\Omega)\right). \label{a26}
\end{align}
The singular terms in $I_2$ can be calculated directly. Hence we have
\begin{align}
\mathcal{C}_{4,c_1}=&\dfrac{L^3}{8 G_N}\left[\dfrac{2\left(1-\cos(\Omega)\right)}{9}\dfrac{H^3}{\delta ^3}-\frac{\cos(\Omega)}{3}\dfrac{ H}{\delta}+\dfrac{\beta(h_0)}{3}\log\left(\dfrac{\delta}{H}\right) \right], \label{a27}
\end{align}
where 
\begin{align}
\beta(h_0)=&2 \int_ {h_0}^{0} dh\big(\dfrac{\sin(\theta)}{h^3\dot{h}}+\dfrac{1}{2} \dfrac{\cos( \Omega)}{h^2}-\frac{1}{8}\cot ^2 (\Omega ) \sin (\Omega ) \csc (2\Omega)(3 - \cos (2\Omega)) \log (h) - \dfrac{1}{8}\cos(\Omega)\cot ^2(\Omega)\big)\nonumber\\ 
&-\dfrac{\cos(\Omega)}{ h_0} -\dfrac{h_0}{4}\cos(\Omega) \cot ^2(\Omega)+\frac{1}{4} h_0( 1-\log(h_0))\cot ^2 (\Omega ) \sin (\Omega ) \csc (2\Omega)(3 - \cos (2\Omega)) .
\end{align}
One can perform similar computations for cones in higher dimensions. We have done this for $c_2$ and $c_3$ in CFT$_5$ and CFT$_6$ respectively. The method is similar to what we have presented in $d=4$, so we will skip the details and report the results in these cases.

In the case of $c_2$ one finds two family of divergent terms proportional to $\log \delta$ and $\log ^2\delta$ which are given by
\begin{align}
\begin{split}
\mathcal{C}_{5,c_2}^{\log}&=\dfrac{L^4 }{8 G_N}\log \left(\frac{\delta}{H}\right)
\left(\int_{h_0} ^0 dh\left[\dfrac{\sin ^2(\theta)}{h^4 \dot{h}}-\dfrac{4 \cos ^2 (\Omega) \cot (\Omega)}{9 h}+\dfrac{2 \cos (\Omega) \sin(\Omega)}{3 h^3}\right] -\dfrac{\cos(\Omega)\sin(\Omega)}{3 h_0 ^2}\right)\nonumber\\
\mathcal{C}_{5,c_2}^{\log ^2}&=\dfrac{L^4}{36 G_N}\cos ^2 (\Omega) \cot (\Omega) \log ^2\left(\dfrac{\delta}{H}\right).
\end{split}
\end{align}
One should note that the $\mathcal{C}_{5,c_2}^{\log ^2}$ is not a universal term.

For the case $c_3$ we find 
\begin{align}
\mathcal{C}_{6,c_3}^{\log }=&\dfrac{L^5\pi}{20 G_N}\log\left(\frac{\delta}{H}\right)\Bigg[\int_ {h_0}^{0} dh\Bigg(\dfrac{\sin ^3(\theta)}{h^5\dot{h}}+\frac{3}{4} \dfrac{\cos(\Omega)\sin ^2(\Omega)}{h^4} - \frac{3}{256}\dfrac{\cos(\Omega)(67+35 \cos(2\Omega))}{h^2}\nonumber\\
 &- \frac{27}{8192}(155+ 106 \cos(2\Omega) +15 \cos(4\Omega)) \cot^2(\Omega) \csc(\Omega)\big)\Bigg) -\dfrac{\cos(\Omega) \sin^2(\Omega)}{4 h_0^3}+\dfrac{3 \cos(\Omega)(67+35\cos(2\Omega)}{256 h_0}\nonumber\\
 &+\frac{27h_0}{8192}(155+ 106 \cos(2\Omega) +15 \cos(4\Omega))\cot^2(\Omega) \csc(\Omega)\big) \Bigg]
\end{align}

%%%%%%%%%%%%%%%%%%%%%%%%%%%%%%%
%%%%%%%%%%%%%%%%%%%%%%%%%%%%%%%
\subsection{Crease $ k\times R^{m} $ }
Consider the following metric for  a $AdS_{d+1}$ space-time in the bulk
\begin{equation}
 ds^2=\dfrac{L^2}{z^2}\left[-d t^2+dz^2+d\rho ^2+ \rho^ 2 d\theta ^2+\sum _{i=1}^{m} (dx ^ i)^{2}\right]. \label{a30}
\end{equation}
 where the Cartesian coordinates $x^i$ denote a $R^m$ flat space for $m=d-3$. Consider a kink subregion defined as  $\theta \in [-\Omega,\Omega]$ and $\rho \in [0,\infty]$ for the full range of $x^i\in[-\infty,\infty]$. However, to avoid IR singularities in the following calculations we restrict ourselves to the limited region $\rho \in [0,H]$ and $x^i \in[-\frac{\tilde{H}}{2},\frac{\tilde{H}}{2}]$. Assume that the extension of the entangling region in the bulk is given by the radial coordinate $z=z(\rho,\theta)$.  
Hence, the induced metric on the extended surface read
\begin{equation}
h=\left(\begin{array}{c c c cc}
\dfrac{L^2}{z^2}(1+(z')^2)&\dfrac{L^2}{z^2}z' \dot{z} &&&\\
\dfrac{L^2}{z^2}z' \dot{z}&\dfrac{L^2}{z^2}(\rho ^2+\dot{z}^2)&&&\\
 &&\dfrac{L^2}{z^2}&&\\
&&&\ddots &\\
&&&&\dfrac{L^2}{z^2} \\ 
\end{array}\right) . \label{a31}
\end{equation} 
The area functional to be minimized is given by
\begin{align}
S_{d,k\times R^{d-3}} =& L^{d-1}\tilde{H}^{d-3}  \int d\rho d \theta\, \dfrac{\sqrt{\dot{z}^2+\rho ^2(1+z'^2)}}{z^{d-1}}. \label{a32}
\end{align}
%and the equation of motion for $ z(\rho,\theta)$ is given by
%\begin{equation}
%\rho z(\rho ^2+\dot{z}^2)z''+\rho z(1+z'^2)\ddot {z} -2\rho z\dot{z}z' \dot{z}'+\dot{z}^2((d-1)\rho +2 zz')+ \rho ^2((d-1)\rho +z z')(1+z'^2)=0. \label{a33}
%\end{equation}
Again one can use the scaling property $z=\rho h(\theta),$ to find the equation of motion as
% Eq. (\ref{a33})
%reduces to
\begin{equation}
h(1+h^2)\ddot{h}+(d-1)\dot{h}^2 +h^4 +d h^2+ (d-1)=0.\label{a34}
\end{equation}
Eq. (\ref{a34}) can be integrated over to find the following constant along the $\theta$ variation
 \begin{equation}
 K_d=\dfrac{(1+h^2)^{\frac{(d-1)}{2}}}{h^{(d-1)} \sqrt{\dot{h}^2+h^2+1}}=\dfrac{(1+ h_0^2)^{\frac{(d-2)}{2}}} {h_0^{(d-1)}}\ .\label{a35}
 \end{equation}
 Noticing that  $h$ is a decreasing function near the boundary, we have from Eq. (\ref{a35})
\begin{equation}
\dot{h}=-\dfrac{ \sqrt{1+h^2}\sqrt{(1+h^2)^{d-2}-K_d^2 h^{2(d-1)}}}{k_d h^{d-1}} \  .\label{a36}
\end{equation}
One can find the volume as
%$V(\gamma)$, using the volume element $(L^d/z^d) d\rho \rho d\theta  \Pi_i dx^i$ for the bulk, we can write
\begin{align}
\begin{split}
V(\gamma)&=L^d \tilde{H}^{d-3}\int \rho d\rho \int d\theta \int _{\delta} ^{z}\dfrac{dz}{z^d}\nonumber\\
&=\dfrac{  L^dH^2\tilde{H}^{d-3} \Omega}{(d-1)\delta ^{d-1}} -\dfrac{2 L^d\tilde{H}^{d-3} }{(d-1)}\int _{\delta / h_0}^H \dfrac{d\rho}{\rho ^{d-2 }} \int _{h_0}^{\delta / \rho}\dfrac{dh}{\dot{h}h^{d-1}}.
\end{split}
\end{align}
In the limit $ h\rightarrow0 $ the integrand in the last term behaves as
\begin{align}
\dfrac{1}{h^{d-1}\dot{h}}\sim - K_d +O(h^2).
\end{align}
So we can write
\begin{align}
\begin{split}
V(\gamma)=&\dfrac{ L^d\tilde{H}^{d-3} \Omega H^2}{(d-1)\delta ^{d-1}}\\
&-\dfrac{2 L^d\tilde{H}^{d-3} }{(d-1)}\int _{\delta / h_0}^H \dfrac{d\rho}{\rho ^{d-2 }} \int _{h_0}^{\delta / \rho}dh \left(\dfrac{1}{\dot{h}h^{d-1}}+ K _d \right) +\dfrac{2 L^d\tilde{H}^{d-3} }{(d-1)}\int _{\delta / h_0}^H \dfrac{d\rho}{\rho ^{d-2 }} \int _{h_0}^{\delta / \rho}dh K _d.
\end{split}
\end{align}
We can separate  the divergent term as follows
\begin{align}
I_1=\int _{\delta / h_0}^H \dfrac{d\rho}{\rho ^{d-2 }} \int _{h_0}^{\delta / \rho}dh \left(\dfrac{1}{\dot{h}h^{d-1}}+ K _d \right).
\end{align}
 Let us denote
\begin{align}
J(h)=\dfrac{1}{\dot{h}h^{d-1}}+ K _d,
\end{align}
it is clear from Eq. (\ref{a36}) that  $J(h) \sim(h^2) $ as $ h\rightarrow0 $. We can find the integral $(I_1)$  by parts 
\begin{align}
I_1=&-\dfrac{1}{(d-3) H^{d-3}}\int_{h_0}^{\delta / H} dh J(h) -\dfrac{\delta}{d-3}\int _{\delta /h_0}^{H}  \dfrac{d\rho}{\rho^{d-1}} J(h)\vert _ {h=\frac{\delta}{\rho}}\nonumber\\
&=-\dfrac{1}{(d-3) H^{d-3}}\int _{h_0}^{\delta / H} dh J(h) -\frac{\delta}{d-3} I_2.
\end{align}
Now for finding  the divergences of $ I_2 $, we make a change of variable from $ \rho$  to $ q=\frac{\delta}{\rho} $ and then Taylor expand the terms around $ \delta=0  $
\begin{align}
I_2=&- \dfrac{1}{\delta^{d-2}}\int _{h_0}^{\delta /H} dq q^{d-3} J(q)\nonumber\\
&=- \dfrac{1}{\delta^{d-2}} \left[\int _{h_0}^0  dq q^{d-3} J(q) + \frac{\delta}{H}\big( q^{d-3} J(q)\big ) _{q={\delta / H}}+\cdots\right] \nonumber\\
&=- \dfrac{1}{\delta^{d-2}} \int _{h_0}^0  dq q^{d-3} J(q) +O(\delta ^d).
\end{align}
From Eq. (\ref{a36})  $ q^{d-3} J(q) \sim q^{d-1} $ for small $q$, hence in the above expression the integral over $ q $  is finite. We have also 
\begin{align}
I_1= \dfrac{\delta}{d-3}\dfrac{1}{\delta^{d-2}} \int _{h_0}^0  dq q^{d-3} J(q)  +finite.
\end{align}
So the singular terms of the volume is as follows
\begin{align}
V(\gamma)=&\dfrac{ L^dH^2\tilde{H}^{d-3} \Omega}{(d-1)\delta ^{d-1}}+\dfrac{2 K_d L^d\tilde{H}^{d-3} }{(d-1)}\left[-\dfrac{h_0}{d-3} (\dfrac{h_0}{\delta})^{d-3}+\dfrac{\delta}{d-2} (\dfrac{h_0}{\delta})^{(d-2)}\right] \nonumber\\
&-\dfrac{2  L^d\tilde{H}^{d-3} }{(d-1)(d-3)\delta ^{d-3}} \int _{h_0}^0  dq q^{d-3} J(q) + \textnormal{finite}.
\end{align}
The complexity is finally given by
\begin{align}
\mathcal{C}_{k\times R^m}=\dfrac{V(\gamma)}{8\pi L G_N}.
\end{align}

\subsection{Conical Crease $ c_{n} \times R^{m}$}
In this section we consider the special cases of $ n=1,2 $ and $ m=1,2 $ in the metric \eqref{a1} which we denote them by cone-crease $ c_{n} \times R^{m}$. As in the previous cases the subregion is restricted to the intervals  $ \theta  \in [0,\Omega] $, $\rho \in [0,H] $ and $x_i \in [-\tilde{H}/2, \tilde{H}/2]$ where $H$ and $\tilde{H}$ indicate IR cut-offs. The extended surface in the bulk is demonstrated by the function $ z=z(\rho,\theta) $ with the following induced metric
\begin{equation}
h=\left(\begin{array}{c c c c c c}
\dfrac{L^2}{z^2}(1+(z')^2)&\dfrac{L^2}{z^2}z' \dot{z} &&&&\\
\dfrac{L^2}{z^2}z' \dot{z}&\dfrac{L^2}{z^2}(\rho ^2+\dot{z}^2)&&&&\\
&&\dfrac{L^2\rho ^2 \sin ^2(\theta)}{z^2 }g_{ab} (S^n)&&&\\
&&&\dfrac{L^2}{z^2}&&\\
&&&&\ddots &\\
&&&&&\dfrac{L^2}{z^2} \\ 
\end{array}\right),
\end{equation} 
 where $ g _{ [a b] }(S^n) $  is the metric of the sphere $ (S^n) $. The surface function to be extrimized is the following
\begin{align}
S_{d,c_n\times R^{m}}=& L^{d-1}\tilde{H}^{m}\Omega _n\int  d\rho  d \theta \dfrac{ \rho ^n \sin ^n(\theta) \sqrt{\dot{z}^2+\rho ^2(1+z'^2)}}{z^{d-1}}.
\end{align} 
The equation of motion for $ z(\rho,\theta)$ after imposing the scaling relation  $z=\rho h(\theta)$ reads
 \begin{align}
  h(1+h^2)\ddot{h}+ n\cot(\theta) h \dot{h}^3+( d+n h^2-& 1) \dot{h}^2+ n\cot(\theta) h (1+h^2)\dot{h}\nonumber\\
&+(n+1) h^4+(d+n) h^2+d-1 =0.
\end{align}
First consider $ n=1 $ and $m=1$, i.e.  $ d=5 $. Let us expand $ y= \sin(\theta) $ and $ \dot{h} $ near the boundary in powers of $h$.
\begin{align}
y=& \sin (\Omega) -\frac{1}{6} h^2 \cos (\Omega) \cot (\Omega) -\frac{1}{432} h^4 (19- 5\cos(2\Omega))\cot ^2(\Omega) \csc(\Omega)+O(h^5),\label{a50} \\
\dot{h} (\theta) =&-\dfrac{3 \tan(\Omega)}{h}+\frac{1}{3}h (8 -\cos(2\Omega)) \csc (2\Omega)+f_0 h^2 \nonumber\\
&-\frac{1}{216}h^3 ( 435 -404 \cos(2\Omega) +52 \cos(4\Omega))\csc ^3(\Omega) \sec(\Omega)+\mathcal{O}(h^4), \label{a51}
\end{align} 
where $ \alpha_0 $ is fixed by the condition $ f( h_0 )=0 $ at $\mathcal{O}\left(h^2\right)$ and vanishes at $\mathcal{O}\left(h^3\right)$.
Using equations (\ref{a50}) and (\ref{a51}) the volume functional is as follows
\begin{align}
V(\gamma)=&\dfrac{L^5 H^3\tilde{H} \Omega _1}{12 \delta ^4}(1- \cos(\Omega))
-\dfrac{L^5 \tilde{H} \Omega _1}{4}\int _{\delta / h_0}^{H} \dfrac{d \rho}{\rho^2 } \int_{h_0}^{\delta /\rho} d h \dfrac{\sin(\theta)}{\dot{h} h^4}\nonumber\\
&=V_1 +V_2.
\end{align}
Near the boundary $h \rightarrow 0$, we have
 \begin{align}
\dfrac{ \sin(\theta)}{\dot{h} h^4} \sim -\dfrac{\cos(\Omega)}{3 h^3}+\dfrac{(-13+5 \cos 2\Omega) \cot (\Omega) \csc(\Omega)}{108 h}-\frac{1}{9} f_0 \cos (\Omega) \cot (\Omega) .
 \end{align}
Now we can use it to make the $h$ integral in holographic complexity finite, i.e.
\begin{align}
V_2=&-\dfrac{L^5 \tilde{H} \Omega _1}{4}\int _{\delta / h_0}^{H} \dfrac{d \rho}{\rho^2 } \int_{h_0}^{\delta /\rho} d h \left[ \dfrac{\sin(\theta)}{\dot{h} h^4}+\dfrac{\cos (\Omega)}{3 h^3}-\dfrac{(-13+5 \cos(2 \Omega) ) \cot (\Omega) \csc (\Omega )}{108 h}\right]\nonumber\\
&-\dfrac{L^5\tilde{H} \Omega _1}{4}\int _{\delta / h_0}^{H} \dfrac{d \rho}{\rho^2 } \int_{h_0}^{\delta /\rho} d h \left[-\dfrac{\cos (\Omega)}{3 h^3}+\dfrac{(-13+5 \cos(2\Omega)) \cot(\Omega) \csc (\Omega)}{108 h}\right]\nonumber\\
&=I_1+I_2,
\end{align}
 where
\begin{align}
\begin{split}
 I_2=&-\dfrac{L^5 \tilde{H} \Omega _1}{4}\Bigg[\dfrac{\cos(\Omega) H}{6 \delta ^2}-\frac{1}{\delta}\left (\dfrac{\cos(\Omega)}{3 h_0}+\dfrac{(-13+5 \cos (2\Omega)) \cot (\Omega) \csc(\Omega) h_0}{108 } \right)\\ 
& \hspace{6cm}-\dfrac{(-13+5 \cos (2\Omega)) \cot (\Omega) \csc(\Omega)}{108 H}\log \left(\frac{\delta}{H}\right)\Bigg] .
\end{split}
 \end{align} 
 Let us indicate the integrand in $ I_1 $ by $ J_5(h) $  and integrate  it by parts   
 \begin{align}
 I_1=&-\dfrac{L^5 \tilde{H} \Omega _1}{4}\int _{\delta / h_0}^{H} \dfrac{d \rho}{\rho^2 } \int_{h_0}^{\delta /\rho} d h J_5(h)\nonumber\\
 &=-\dfrac{L^5 \tilde{H} \Omega _1}{4}\left[-\frac{1}{H}\int_{h_0}^{\delta / H} d h J_5(h) - \delta \int _{\delta / h_0}^{H} \dfrac{d \rho}{\rho^3 }J_5(h)\vert _ {h=\delta / \rho}\right].
\label{m38}
 \end{align}
Near the boundary $ J_5(h) \sim \mathcal{O}(h^0) $ . We further make the coordinate transformation $ q=\delta/ \rho $ and Taylor expand the second term of Eq. (\ref{m38}) in terms of $\delta $
\begin{align}
I_1=&-\dfrac{L^5 \tilde{H} \Omega _1}{4}\left[-\frac{1}{H}\int _{h_0}^{\delta / H} d h J_5 (h)+1/\delta \int _ {h_0}^{\delta / H} dq  q  J_5(q)\right] \nonumber\\
&=-\dfrac{L^5 \tilde{H} \Omega _1}{4} \left[-\frac{1}{H}\int _{h_0}^{0} d h J_5 (h)+1/\delta \int _ {h_0}^{0} dq  q  J_5(q) -\dfrac{\delta f_0 \cos(\Omega) \cot(\Omega)}{ 9 H^2}\right]+\mathcal{O}(\delta) .
\end{align}
So we have
\begin{align}
\begin{split}
V(\gamma)=&-\dfrac{L^5 \tilde{H} \Omega _1}{4}\Bigg[-\dfrac{H^3}{3 \delta ^4}( \cos(\Omega)-1)+\dfrac{\cos(\Omega) H}{6 \delta ^2}
\\&
-\frac{1}{\delta}\left(\dfrac{\cos(\Omega)}{3 h_0}+\dfrac{(-13+5 \cos (2\Omega)) \cot (\Omega) \csc(\Omega) h_0}{108 }-\int _ {h_0}^{0} dq  q  J_5(q) \right)\\ &-\dfrac{(-13+5 \cos (2\Omega)) \cot (\Omega) \csc(\Omega) }{108 H}\log \left(\frac{\delta}{H}\right)\Bigg]+ \textnormal{finite}
\end{split}
\end{align}
For $ c_1 \times R^2 $ the result is as follows
\begin{align}
V(\gamma )=& -\dfrac{L ^6 \Omega _1 \tilde{H}^2}{5}\left[ -\dfrac{H^3(1- \cos(\Omega))}{3 \delta ^5}
+\dfrac{\cos (\Omega) H}{12 \delta ^3} \right.\nonumber\\
&+\dfrac{1}{\delta ^2} \left( \dfrac{-\cos(\Omega)}{8 h_0}-\dfrac{h_0 \cot (\Omega) \csc (\Omega)(-11+5\cos(2\Omega))}{512}+\frac{1}{2}\int _{h_0} ^{0} dq q^2 J(q)\right)\nonumber\\
& \left. +\dfrac{ \cot (\Omega) \csc (\Omega)(-11+5\cos(2\Omega))}{256 H \delta} \right] ,
\end{align}
where
\begin{align}
J(h)=\dfrac{\sin (\theta)}{ h^5 \dot {h}}+\dfrac{\cos(\Omega)}{4 h^4}-\dfrac{\cot(\Omega)\csc(\Omega) (-11+5 \cos(2\Omega))}{256 h^2}.
\end{align}
For $ c_2\times R ^1 $ similar steps leads to
\begin{align}
\begin{split}
V(\gamma)=&-\dfrac{\Omega _2  L^6 \tilde{H} }{5}\Bigg[-\dfrac{H^4(\Omega -\frac{1}{2} \sin (2\Omega))}{8 \delta ^5}+\dfrac{\cos (\Omega) \sin (\Omega) H^2}{12 \delta ^3 }\\
&+\frac{1}{\delta}\left(-\dfrac{\cos (\Omega)\sin(\Omega)}{4 h_0^2}-\dfrac{\cos ^2(\Omega) \cot (\Omega) (\log (h_0)-1)}{16}+\int _{h_0} ^{0 }dq q J(q) \right)\\
&+\dfrac{\cos ^2(\Omega) \cot (\Omega)}{16}\dfrac{1}{\delta}\log\left(\frac{\delta}{H}\right)\Bigg],
\end{split}
\end{align}
where
\begin{align}
J(h)=\dfrac{\sin ^2(\theta)}{\dot{h} h^5}+\dfrac{\cos(\Omega) \sin (\Omega)}{2 h^4}-\dfrac{\cos ^2(\Omega) \cot (\Omega)}{16 h^2}.
\end{align}

%%%%%%%%%%%%%%%%%%%%%%%%%%%%%%%%%
%%%%%%%%%%%%%%%%%%%%%%%%%%%%%%%%%
\section{Curved Locus Singular Surfaces} 
In this section, we consider several singular embeddings which have curved locus such as $ k\times\Sigma $ and $ c_n \times\Sigma $ , where locus $ \Sigma $ will take the form $ S^m $ or $ S ^{m-p} \times R^p $.
\subsection{Crease $ k \times \Sigma $}
Consider the geometries $ k\times S^2 $ , $ k\times S^3 $ and $ k\times R \times S^2 $ . We will see that singularities with even dimensional locus will contribute through a logarithmic term.
To begin with, let us consider $ d=5 $ CFT on background $ R^3\times S^2 $ . The action for six-dimensional dual Einstein gravity reads
\begin{align}
I_6 =\frac{1}{l_p ^4} \int d^6 x \sqrt{- g} \left[\frac{20}{L^2}+ R \right].
\label{e25}
\end{align}
We consider the following ansatz for the solution,
\begin{align}
d s^2= \frac{L^2}{z^2}\big [dz^2 +f_1 (z) (dt^2 +d\rho ^2+ \rho^2 d \theta ^2 )+f_2 (z) R_1^2 d \Omega _2 ^2\big ],
\label{e27}
\end{align}
where $ d \Omega_2 ^2 =d \xi_0^2 +\sin ^2 (\xi _0) d \xi_1^2 $ represents a two-sphere metric and $ f_1 $ and $ f_2 $ are functions of the radial coordinate. The boundary of this solution is $ R^3 \times S^2 $ with $ R_1 $ the radius of $ S^2$; so we can recover the flat boundary results in the limit $ R_1 \rightarrow\infty $. Using the Fefferman-Graham expansion near the boundary to find $f_1$ and $f_2$ leads to
\begin{align}
f_1 &= 1+\dfrac{z^2}{12 R_1^2}+ \dfrac{17 z^4}{576 R_1^4}-\dfrac{z^6}{324 R_1^6}+\cdots\\
f_2 &= 1-\dfrac{z^2}{4 R_1^2}-\dfrac{5 z^4}{192 R_1^4}+\dfrac{z^6}{72 R_1^6}+\cdots
\label{e28}
\end{align}
The subregion of interest here is $ \rho\in [ 0, H ] $ and $ \theta \in [-\Omega , \Omega] $  where $ H $ is again a IR cut-off. The coordinates are $ (z , \theta ,\xi_0 , \xi_1 ) $ on the minimal surface  and $ \rho =\rho(z, \theta) $ on the sphere. 
In the limit $ R_1\rightarrow \infty $ one may expect from the case of entanglement entropy that leading order correction to the holographic subregion complexity would be $\mathcal{O}(1/R_1^2) $, however, we show that in this case there is no new divergent term up to $\mathcal{O}(1/R_1^4)$. We first work out the solution $\rho(z,\theta)$ in this approximation with the following ansatz
\begin{align}
\rho(z,\theta)=\dfrac{z}{h(\theta)} +\dfrac{z^2}{R_1} g_2 (\theta)+\dfrac{z^3}{R_1^2}g_3 (\theta)+\dfrac{z^5}{R_1^4}g_5 (\theta)+\mathcal{O}\left(z^7\right)
\label{e33}
\end{align} 
Using the ansatz (\ref{e33}) in the equation of motion of $ \rho(z,\theta) $ leads to vanishing of even terms $ g_{2n}$.
In order to separate the logarithmic divergence, we impose $ \rho= \rho_0 (z,\theta)+\rho_1(z,\theta)/R_1^2+\rho_2(z,\theta)/R_1^4 $, where $ \rho_0=z/h(\theta) $ and $ \rho_1=z^3 g_3 (\theta)$, $\rho_2=z^5 g_5(\theta) $ are higher corrections in the large $ R_1 $ regime.
Now we come back to the metric (\ref{e27}) and find the volume holographic complexity as
\begin{align}
V(\gamma)= &L^5 R _1 ^2 \Omega _ 2 \int d\rho d\theta dz \dfrac{f_1 f_2 \rho}{z^5}\nonumber\\
&= L^5 R_{1}^{2} \Omega _2 \int d \theta dz \dfrac{f_1 f_2 }{z^5} \int _{\rho(z,\theta)}^{H} d\rho \rho \nonumber\\
&=\dfrac{ L^5 R_{1}^{2} \Omega _2}{2} \left(- \int d \theta dz \dfrac{f_1 f_2 \rho ^2}{z^5}+ H^2 \int d \theta dz \dfrac{f_1 f_2 }{z^5}\right) \nonumber\\
&=V_1+V_2
\label{e34}
\end{align}
Now, we can insert the ansatz $ \rho= \rho_0+\rho_1/R_1 ^2 +\rho_2/R_1 ^4 $ that $ \rho_0=z/h(\theta) $, $ \rho_1=z^3 g_3(\theta) $ and $\rho_2=z^5 g_5(\theta)$ and use (\ref{e28}) in the integrand to simplify the results as
\begin{align}
\begin{split}
V_1=&L^5 R _1 ^2 \Omega _ 2 \Bigg[\int _{z_m}^{\delta}\frac{dz}{z^3} \int _{h_0}^{h_1c}\dfrac{dh}{\dot{h}h^2}-\frac{1}{6 R_1^2}\int _{z_m}^{\delta} \frac{dz}{z}\int _{h_0}^{h_1c}\dfrac{dh}{\dot{h}h^2}+\frac{2}{R_1 ^2}\int _{z_m}^{\delta} \frac{dz}{z}\int _{h_0}^{h_1c}\dfrac{dh g_3(\theta)}{\dot{h}h}\\
&-\dfrac{5}{288 R_1^4}\int _{z_m}^{\delta }dz z \int _{h_0}^{h_1c}\dfrac{dh}{\dot{h}h^2}-\frac{1}{3 R_1^4}\int _{z_m}^{\delta} dz z\int _{h_0}^{h_1c}dh\dfrac{g_3}{\dot{h}h}+\frac{1}{R_1 ^4}\int _{z_m}^{\delta} dz z\int _{h_0}^{h_1c}dh\dfrac{g_3^2(\theta)}{\dot{h}}\\
&+\frac{2}{R_1 ^4}\int _{z_m}^{\delta} dz z\int _{h_0}^{h_1c} dh\dfrac{g_5(\theta)}{\dot{h}h}\Bigg],
\end{split}
\label{e35}
\end{align}
and
\begin{align}
V_2=L^5 R _1 ^2 \Omega _ 2 H^2 \Omega \left(\frac{1}{4\delta ^4}-\dfrac{1}{12 R_1^2 \delta ^2}\right)
&+\dfrac{5 \Omega L^5 R _1 ^2 \Omega _ 2 H^2}{288 R_1^4}\log \delta
+\mathrm{finite}
\label{e36}
\end{align}
where $ \delta $ is the UV cut-of.  We have also  changed the integration limits from $ (-\Omega,\Omega) $ to $ (0,\Omega) $ and then changed the integration variable in $ V_1 $ to $ h(\theta) $.
It is instructive to use the following constant of motion
\begin{align}
K_5= \dfrac{(1+h^2)^2}{h^4 \sqrt{1+h^2 +\dot{h}^2}},
\label{e37}
\end{align} 
which is related to $ h(0) $ at the turning point. To find the logarithmic divergent parts it is enough to find the asymptotic behavior of $ h $ and $ g_3 $. Solving $ g_3 $ in terms of $ h $ in the limit of small $ h $ leads to
\begin{align}
g_3=\frac{b_3}{h^3}+\dfrac{1+88 b_3}{56 h}+\dfrac{4+72 b_3}{189}h+\mathcal{O}\left(h^3\right),
\label{e38}
\end{align}
\begin{align}
g_5=\frac{9 b_3^2}{5h^5}+\dfrac{b_3(345+15856b_3)}{7000 h^3}+\mathcal{O}\left(h^{-1}\right),
\label{k38}
\end{align}
where $ b_3 $ can be fixed by demanding $ g_3 $ to have an extremum at $\theta=0$.
We will need to find the series expansion of $ h_{1c} $ in terms of $ \delta $ as follows
\begin{align}
h_{1c}(\delta)= \left(\frac{1}{H}+\dfrac{b_3 H}{R_1^2}-\dfrac{b_3^2 H^3}{5 R_1^4}\right) \delta +\Bigg(\dfrac{(1+88 b_3)}{56 H R_1^2}+\dfrac{(1+88 b_3)H b_3}{56  R_1^4}\Bigg) \delta ^3+ \mathcal{O}\left(\delta ^5\right)
\label{e39}
\end{align}
where $ h_{1c}=h(\Omega-\epsilon) $. The result is obtained for the leading corrections in $ R_1 $ at any order of $ \delta$ \cite{ref.1}.
Now we look at (\ref{e35}) to analyze the divergent terms in the asymptotic limit
\begin{align}
\dfrac{1}{\dot{h} h^2}\sim - K_5 h^2 +2 K_5 h^4+\mathcal{O}\left(h^6\right)
\label{e40}
\end{align}
and
\begin{align}
\dfrac{g_3}{\dot{h} h}\sim - K_5 b_3 -\dfrac{K_5(1-24 b_3)h^2}{56}+\mathcal{O}\left(h^4\right).
\label{e41}
\end{align}
\begin{align} 
\dfrac{g_3^2}{\dot{h} }\sim -\dfrac{K_5 b_3^2}{h^2}-\dfrac{k_5(b_3+32b_3^2)}{28}+\mathcal{O}\left(h^2\right).
\label{g42}
\end{align}
\begin{align}
\dfrac{g_5}{\dot{h} h}\sim -\dfrac{9 K_5 b_3^2}{5h^2}- \dfrac{b_3(345-9344b_3)k_5}{7000}+\mathcal{O}\left(h^2\right).
\label{g43}
\end{align}
We organiz different terms of the integrand in following form
\begin{align}
I_1=\int _{z_m}^{\delta }\frac{dz}{z^3} \int _{h_0}^{h_1c}\dfrac{dh}{\dot{h}h^2}
\label{e42}
\end{align}
\begin{align}
I_2=\int _{z_m}^{\delta }\frac{dz}{z} \int _{h_0}^{h_1c}\dfrac{dh}{\dot{h}h^2}
\label{e43}
\end{align}
\begin{align}
I_3=&\int _{z_m}^{\delta }\frac{dz}{z} \int _{h_0}^{h_1c}\dfrac{dh g_3}{\dot{h}h}=\int _{z_m}^{\delta }\frac{dz }{z} \int _{h_0}^{h_1c} dh\left(\dfrac{ g_3}{\dot{h}h}+K_5 b_3 \right)-K_5 b_3\int _{z_m}^{\delta} \frac{dz }{z}(h_{1c}-h_0)\nonumber\\
&=I'_1+I'_2
\label{e44}
\end{align}
\begin{align} 
I''_1=\int _{z_m}^{\delta } z dz \int _{h_0}^{h_1c} \dfrac{dh}{\dot{h}h^2}
\label{e43}
\end{align}
\begin{align}
I''_2=\int _{z_m}^{\delta } z dz \int _{h_0}^{h_1c} dh\dfrac{g_3}{\dot{h}h^2}
\label{e43}
\end{align}
\begin{align}
I''_3=&\int _{z_m}^{\delta } z dz \int _{h_0}^{h_1c}dh(\dfrac{g_3^2}{\dot{h}}+\dfrac{k_5 b_3^2}{h^2})-\int _{z_m}^{\delta } z dz \int _{h_0}^{h_1c}dh\dfrac{k_5 b_3^2}{h^2}\nonumber\\
&=I'''_1+I'''_2
\label{e43}
\end{align}
\begin{align}
I''_4=&\int _{z_m}^{\delta } z dz \int _{h_0}^{h_1c} dh(\dfrac{g_5}{\dot{h}h}+\dfrac{9k_5 b_3^2}{5h^2})-\int _{z_m}^{\delta } z dz \int _{h_0}^{h_1c} dh\dfrac{9k_5b_3^2}{5h^2}\nonumber\\
&=I'''_3+I'''_4
\label{e43}
\end{align}
Now we differentiate each of them with respect to the UV cut-off and look for $ 1/\delta $ divergent terms. One can easily find
\begin{align}
\dfrac{d I_1}{d\delta}=&\frac{1}{\delta ^3} \int _{h_0}^{h_{1c}}\dfrac{dh}{\dot{h}h^2}\nonumber\\
&=\frac{1}{\delta ^3} \int _{h_0}^{0}\dfrac{dh}{\dot{h}h^2}+\frac{1}{\delta ^2}\frac{d h_{1c}}{d\delta}\left[\dfrac{1}{\dot{h}h^2}\right]_{h=h_{1c}}+\cdots\nonumber\\
&=\frac{1}{\delta ^3} \int _{h_0}^{0}\dfrac{dh}{\dot{h}h^2}+\mathcal{O}\left(\delta ^0\right).
\label{e45}
\end{align}
\begin{align}
\dfrac{d I_2}{d\delta}=&\frac{1}{\delta} \int _{h_0}^{h_{1c}}\dfrac{dh}{\dot{h}h^2}\nonumber\\
&=\frac{1}{\delta} \int _{h_0}^{0}\dfrac{dh}{\dot{h}h^2}+\frac{d h_{1c}}{d\delta}\left[\dfrac{1}{\dot{h}h^2}\right]_{h=h_{1c}}+\cdots\nonumber\\
&=\frac{1}{\delta} \int _{h_0}^{0}\dfrac{dh}{\dot{h}h^2}+\mathcal{O}\left(\delta ^2\right).
\label{e46}
\end{align}
\begin{align}
\frac{d I'_1}{d\delta}=&\frac{1}{\delta} \int _{h_0}^{h_1c} dh\big[\dfrac{ g_3}{\dot{h}h}+K_5 b_3\big]\nonumber\\
&=\frac{1}{\delta} \int _{h_0}^{0} dh\left(\dfrac{ g_3}{\dot{h}h}+K_5 b_3\right)+\frac{d h_{1c}}{d\delta}\left[\dfrac{g_3}{\dot{h}h}+K_5 b_3\right]_{h=h_{1c}}+...\nonumber\\
&=\frac{1}{\delta} \int _{h_0}^{0} dh\left(\dfrac{ g_3}{\dot{h}h}+K_5 b_3\right)+\mathcal{O}\left(\delta ^2\right).
\label{e47}
\end{align}
\begin{align}
\frac{d I'_2}{d\delta}=&-\frac{k_5 b_3}{\delta}(h_{1c}-h_0)\nonumber\\
&=\frac{ h_0k_5 b_3}{\delta}-\dfrac{k_5 b_3}{H}(1+b_3 H^2/R_1^2-b_3^2 H^4/{5 R_1^4})+\mathcal{O}\left(\delta ^2\right).
\label{e48}
\end {align}
\begin{align} 
\dfrac{d I''_1}{d\delta}=&\delta \int _{h_0}^{h_{1c}}\dfrac{dh}{\dot{h}h^2}\nonumber\\
&=\delta \int _{h_0}^{0}\dfrac{dh}{\dot{h}h^2}+\delta ^2\frac{d h_{1c}}{d\delta}\left[\dfrac{1}{\dot{h}h^2}\right]_{h=h_{1c}}+\cdots\nonumber\\
&=\delta \int _{h_0}^{0}\dfrac{dh}{\dot{h}h^2}+\mathcal{O}\left(\delta ^4\right).
\label{g42}
\end{align}
\begin{align}
\dfrac{d I''_2}{d\delta}=&\delta \int _{h_0}^{h_{1c}} dh\dfrac{g_3}{\dot{h}h}\nonumber\\
&=\delta \int _{h_0}^{0} dh\dfrac{g_3}{\dot{h}h^2}+\delta ^2\frac{d h_{1c}}{d\delta}\left[\dfrac{g_3}{\dot{h}h}\right]_{h=h_{1c}}+\cdots\nonumber\\
&=\delta \int _{h_0}^{0}\dfrac{dh}{\dot{h}h^2}+\mathcal{O}\left(\delta ^2\right).
\label{g43}
\end{align}
\begin{align}
\dfrac{d I'''_1}{d\delta}=&\delta \int _{h_0}^{h_{1c}} dh\left(\dfrac{g_3 ^2}{\dot{h}}+\dfrac{k_5 b_3^2}{h^2}\right)\nonumber\\
&=\delta \int _{h_0}^{0} dh\left(\dfrac{g_3^2}{\dot{h}}+\dfrac{k_5 b_3^2}{h^2}\right)+\delta ^2\frac{d h_{1c}}{d\delta}\left[\dfrac{g_3^2}{\dot{h}}+\dfrac{k_5 b_3^2}{h^2}\right]_{h=h_{1c}}+\cdots\nonumber\\
&=\delta \int _{h_0}^{0}dh\left(\dfrac{g_3^2}{\dot{h}}+\dfrac{k_5 b_3^2}{h^2}\right)+\mathcal{O}\left(\delta ^2\right).
\label{g44}
\end{align}
\begin{align}
\dfrac{d I'''_2}{d\delta}=&-\delta \int _{h_0}^{h_{1c}}\dfrac{k_5 b_3^2}{h^2}\nonumber\\
&=\delta k_5 b_3^2\left(\dfrac{1}{h_{1c}}-\dfrac{1}{h_0}\right)\nonumber\\
&=-\dfrac{\delta k_5 b_3^2}{h_0}+k_5 b_3^2H\left[1-\dfrac{b_3 H^2}{R_1^2}+\frac{b_3^2 H^4}{5R_1^4}\right]+\mathcal{O}\left(\delta ^2\right).
\label{g45}
\end{align}
\begin{align}
\dfrac{d I'''_3}{d\delta}=&\delta \int _{h_0}^{h_{1c}} dh\left(\dfrac{g_5}{\dot{h}h}+\dfrac{9k_5 b_3^2 }{5h^2}\right)\nonumber\\
&=\delta \int _{h_0}^{0} dh\left(\dfrac{g_5}{\dot{h}h}+\dfrac{9k_5 b_3^2}{5h^2}\right)+\delta ^2\frac{d h_{1c}}{d\delta}\left[\dfrac{g_5}{\dot{h}h}+\dfrac{9k_5 b_3^2 }{5h^2}\right]_{h=h_{1c}}+\cdots\nonumber\\
&=\delta \int _{h_0}^{0}dh\left(\dfrac{g_5}{\dot{h}h}+\dfrac{9k_5 b_3^2}{5h^2}\right)+\mathcal{O}\left(\delta ^2\right).
\label{g45}
\end{align}
\begin{align}
\dfrac{d I'''_4}{d\delta}=&-\delta \int _{h_0}^{h_{1c}}\dfrac{9k_5 b_3^2}{5h^2}\nonumber\\
&=\delta \frac{9k_5 b_3^2}{5}\left(\dfrac{1}{h_{1c}}-\dfrac{1}{h_0}\right)\nonumber\\
&=-\delta \dfrac{9 k_5 b_3^2}{5h_0}+\frac{9k_5 b_3^2H}{5}\left[1-\dfrac{b_3 H^2}{R_1^2}+\frac{b_3^2 H^4}{5R_1^4}\right]+\mathcal{O}\left(\delta ^2\right).
\label{g46}
\end{align}
So from (\ref{e42})-(\ref{e48}) we can find the logarithmic divergences in the holographic complexity for $ k\times S^2 $ geometry as follows
\begin{align}
\mathcal{C}^{\log}_{k\times S^2}=\dfrac{L^5 R_1^2 \Omega _2}{8 \pi L G} \left[-\frac{1}{6 R_1^2}\int _{h_0} ^0 \frac{dh}{\dot{h} h^2}+\frac{2}{R_1^2} \int _{h_0} ^0 dh\left(\frac{g_3}{\dot{h} h}+k_5 b_3\right) +\dfrac{2 h_0 k_5 b_3}{R_1^2}+\dfrac{5 \Omega L^5 R _1 ^2 \Omega _ 2 H^2}{288 R_1^4}\right] \log (\delta)
\label{g48}
\end{align}
Note that in this case no new divergent term appears due to the singular surface. All new $\log\delta$ terms are suppressed with a factor of $\delta^\alpha$ where $\alpha\ge 1$. 
\subsubsection*{Subregion $ k\times S^3 $}
Now we want to find the holographic subregion complexity for $ k\times S^3 $ geometry in a CFT on $R^3\times S^3 $. We will show that in this case the singularity gives no logarithmic contribution to subregion complexity. Consider the following metric 
\begin{align}
d s^2= \frac{L^2}{z^2}\big [dz^2 +f_1 (z) (dt^2 +d\rho ^2+ \rho^2 d \theta ^2 )+f_2 (z) R_1^2 d \Omega _3 ^2\big ],
\label{e50}
\end{align}
where $ d \Omega_3 ^2 =d \xi_0^2 +\sin ^2 (\xi _0) d \xi_1^2+ \sin ^2 (\xi _0)\sin ^2 (\xi _1) d \xi_2 ^2 $ is the unit $S^3$ and we find $ f_1 $ and $ f_2 $ as
\begin{align}
\begin{split}
f_1 &= 1+\dfrac{3 z^2}{20 R_1^2}+ \dfrac{69 z^4}{1600 R_1^4}+ \dfrac{z^6}{R_1^6}\left(\frac{33}{8000}-\frac{1}{200}\log{z}\right)+\mathcal{O}\left(z^8\right)\\
f_2&= 1-\dfrac{ 7 z^2}{20 R_1^2}-\dfrac{11 z^4}{1600 R_1^4}+ \dfrac{z^6}{R_1^6}\left(\frac{67}{8000}+\frac{1}{200}\log{z}\right)+\mathcal{O}\left(z^8\right)
\label{e51}
\end{split}
\end{align}
Similar to the previous case the induced coordinates on the RT surface are $ (z, \theta ,\xi_0 ,\xi_1 , \xi_2) $ and $\rho=\rho(z,\theta)$. Using the equation of motion for $ h $ we can find the following constant of motion 
\begin{align}
K_6= \dfrac{(1+h^2)^{5/2}}{h^5 \sqrt{1+h^2 +\dot{h}^2}},
\label{e55}
\end{align} 
which can be fixed in terms of the boundary data. Using the metric (\ref{e50}) we find the holographic complexity as
\begin{align}
V(\gamma)= &L^6 R _1 ^3 \Omega _ 3\int d\rho d\theta dz \dfrac{f_1 f_2 ^{3/2} \rho}{z^6}\nonumber\\
&= L^6 R_{1}^{3} \Omega _3 \int d \theta dz \dfrac{f_1 f_2 ^{3/2}}{z^6} \int _{\rho(z,\theta)}^{H} d\rho \rho \nonumber\\
&=\dfrac{ L^6 R_{1}^{3} \Omega _3}{2} \left(- \int d \theta dz \dfrac{f_1 f_2 ^{3/2} \rho ^2}{z^6}+ H^2 \int d \theta dz \dfrac{f_1 f_2^{3/2 }}{z^6}\right) \nonumber\\
&=V_1+V_2
\label{e56}
\end{align}
 Inserting the ansatz $ \rho=\rho_0 +\rho_1/R_1^2 $, $ \rho_0=z/h(\theta) $and $ \rho_1=z^3 g_3(\theta) $ and using the expansions (\ref{e51}) in the integrand, simplifies the result as 
\begin{align}
V_1=L^6 R _1 ^3 \Omega _ 3\int _{z_m}^\delta\frac{dz}{z^2}\int _{h_0}^{h_1c}\dfrac{dh}{\dot{h}h^2}\left(\frac{1}{2z^2}-\frac{9}{20 R_1^2}+\frac{1}{R_1 ^2}hg_3(\theta)\right),
\label{e57}
\end{align}
and
\begin{align}
V_2=L^6 R _1 ^3 \Omega _ 3 H^2 \Omega \left(\frac{1}{10\delta ^5}-\dfrac{9}{60 R_1^2 \delta ^3}\right)+\mathrm{finite},
\label{e58}
\end{align}
where $ \delta $ is the UV cut-of, such that $ \rho(z, \Omega-\epsilon) = H $ and $ z_m $ is defined such that $ \rho(z_m,0)=H $. We have also changed the integration limits from $ (-\Omega,\Omega) $ to $ (0,\Omega) $ and then changed the integration variable in $ V_1 $ to $ h(\theta) $.

Similar to what we have done in the previous sections in details, one can work out the logarithmic divergence in this case. Here we step the details and report to the final result
\begin{align}
\mathcal{C}_{k\times S^3}=\dfrac{L^6 R_1^3 \Omega _3}{8 \pi L G_N} \left[\dfrac{\Omega H^2 }{10 \delta ^5} -\dfrac{9 \Omega H^2}{60 R_1 ^2 \delta ^3}-\frac{1}{6 \delta ^3}\int _{h_0} ^0 \frac{dh}{\dot{h} h^2}+\frac{9}{ 20 R_1^2 \delta} \int _{h_0} ^0 \frac{dh}{\dot{h} h^2}-\frac{1}{ R_1^2 \delta}\int _{h_0} ^0 dh\frac{g_3}{\dot{h} h} \right]. 
\label{e69}
\end{align}
In this case no new divergent term appears due to the singular surface and all new $\log\delta$ terms are suppressed with a factor of $\delta^\alpha$ where $\alpha\ge 1$. 
\subsubsection*{Subregion $ k\times R^1 \times S^2 $}
In the following we give another example showing that odd dimensional locus does not contribute to logarithmic singularities, althogh it has non-zero curvature. We consider a CFT defined on $ R^4 \times S^2 $. The bulk metric is given by
\begin{align}
d s^2= \frac{L^2}{z^2}\big [dz^2 +f_1 (z) (dt^2 +d\rho ^2+ \rho^2 d \theta ^2+d x^2 )+f_2 (z) R_1^2 d \Omega _2 ^2\big ],
\label{e70}
\end{align}
where $ d\Omega _2 $ is the line element over $ S^2 $ and $ f_1$ and  $ f_2 $ have the following expansions
\begin{align}
f_1 = 1+\dfrac{ z^2}{20 R_1^2}+ \dfrac{ z^4}{100 R_1^4}+\dfrac{z^6}{ R_1^6}\left(\frac{1}{1200}-\frac{1}{400}\log{z}\right)+\mathcal{O}\left(z^8\right),\nonumber\\
f_2= 1-\dfrac{z^2}{5 R_1^2}-\dfrac{7 z^4}{800 R_1^4}+\dfrac{z^6}{ R_1^6}\left(\frac{7}{4800}+\frac{1}{200}\log{z}\right)+\mathcal{O}\left(z^8\right).
\label{e71}
\end{align}
The subregion $ k \times R ^1 \times S^2 $ is defined by $ \theta \in [-\Omega,\Omega], \ x \in[-\infty ,\infty] $ and $ \rho \in [0,\infty] $. we put IR cut-offs on $ x $ and $ \rho $ directions such that $ x\in [-\tilde{H}/2 , \tilde{H}/2 ] $ and $ \rho \in[\rho_m , H] $, where $\rho_m $ is given in terms of $\delta $. Similar to the previous cases $ (z, \theta, x,\xi_0, \xi_1) $ are the coordinates on the RT surface with $ \rho =\rho (z,\theta)$. The equation of motion for $ h $ gives the following constant of motion
\begin{align}
K_6= \dfrac{(1+h^2)^{5/2}}{h^5 \sqrt{1+h^2 +\dot{h}^2}},
\label{e75}
\end{align} 
Returning to the metric (\ref{e70}) we find the holographic complexity as
\begin{align}
V(\gamma)= &L^6 R _1 ^2 \tilde{H}\Omega _ 2\int d\rho d\theta dz \dfrac{f_2 f_1 ^{3/2} \rho}{z^6}\nonumber\\
&= L^6 R_{1}^{2}\tilde{H} \Omega _2\int d \theta dz \dfrac{f_2 f_1^{3/2}}{z^6} \int _{\rho(z,\theta)}^{H} d\rho \rho \nonumber\\
&=\dfrac{ L^6 R_{1}^{2}\tilde{H} \Omega _2}{2} \left(- \int d \theta dz \dfrac{f_2 f_1 ^{3/2} \rho ^2}{z^6}+ H^2 \int d \theta dz \dfrac{f_2 f_1^{3/2 }}{z^6}\right) \nonumber\\
&=V_1+V_2.
\label{e76}
\end{align}
We then insert the ansatz $ \rho=\rho_0+\rho_1/R_1^2 $, $ \rho_0=z/h(\theta)$ and $ \rho_1=z^3 g_3(\theta) $ and use (\ref{e71}) in the integrand to simplify the expressions as follows
\begin{align}
V_1=-L^6 R _1 ^2 \tilde{H}\Omega _ 2 \int^{z_m}_\delta \frac{dz}{z^2}\int _{h_0}^{h_1c}\dfrac{dh}{\dot{h}h^2}\left(\frac{1}{z^2} -\frac{1}{8 R_1^2}+\frac{2}{R_1 ^2}h g_3(\theta)\right),
\label{e77}
\end{align}
and
\begin{align}
V_2=L^6 R _1 ^2\Omega _ 2 H^{2} \tilde{H} \Omega \left(\frac{1}{5\delta ^5}-\dfrac{1}{24 R_1^2 \delta ^3}\right)+\mathrm{finite},
\label{e78}
\end{align}
Again we step the details of the rest of this calculation we find 
\begin{align}
\mathcal{C}_{k \times R^1\times S^1}=\dfrac{L^6 R_1^2 \tilde{H}\Omega _2}{8 \pi L G} \left(\dfrac{\Omega H^2 }{5 \delta ^5}-\dfrac{\Omega H^2}{24 R_1 ^2 \delta ^3}-\frac{1}{3 \delta ^3}\int _{h_0} ^0 \frac{dh}{\dot{h} h^2}+\frac{1}{ 8 R_1^2 \delta} \int _{h_0} ^0 \frac{dh}{\dot{h} h^2}-\frac{2}{ R_1^2 \delta}\int _{h_0} ^0 dh\frac{g_3}{\dot{h} h}\right). 
\label{e88}
\end{align}
As the case of $k\times S^2$ and $k\times S^3$ new logarithmic divergent term in this case are also suppressed with a factor of $\delta^\alpha$ with a positive power. 

%%%%%%%%%%%%%%%%%%%%%%%%%%%%%
\subsection{Conical Crease $ c_n \times \Sigma $}
In this section, we will calculate holographic complexity for subregions with conical singularities of the form $ c_n\times S^m$.

\subsubsection*{Subregion $c_1\times S^1$}
To begin with, we concider the simplest case with $m=1$. In this case, the background geometry for CFT is $ R^4\times S^1$. The dual bulk geometry is then given by 
\begin{align}
d s^2= \frac{L^2}{z^2}\big [dz^2 +f_1 (z) (dt^2 +d\rho ^2+ \rho^2 d \theta ^2 +\rho ^2 \sin ^2(\theta) d \phi ^2)+f_2 (z) R_1^2 d \xi _0^2\big ],
\label{f_1}
\end{align}
where $ f_1=1+O(1/R_1^6) $ and $ f_2=1+O(1/R_1^6) $. The singular subregion of our interest is defined as $\theta \in [0,\Omega], \ \xi_0 \in[0 ,2 \pi] $, $ \phi \in[0 ,2 \pi] $ and $ \rho \in [0, H] $. 

One can find that $ g_3 =0 $ is the exact solution for this case and since the equation of motion for $ h $ is the same as $ c_1 \times R^1$ case, the holographic subregion complexity might become same. Returning to the metric (\ref{f_1}) gives the holographic complexity as
\begin{align}
V(\gamma)= &L^5 R _1 4\pi ^2 \int d\rho d\theta dz \dfrac{f_2 ^{1/2} f_1 ^{3/2} \rho ^2 \sin (\theta)}{z^5}\nonumber\\
&=L^5 R _1 4\pi ^2\int d \theta dz \dfrac{f_2 ^{1/2} f_1^{3/2}\sin (\theta)}{z^5} \int _{\rho(z,\theta)}^{H} d\rho \rho ^2 \nonumber\\
&=\dfrac{L^5 R _1 4\pi ^2}{3}\left(- \int d \theta dz \dfrac{f_2 ^{1/2}f_1 ^{3/2} \rho ^3}{z^5}+ H^3 \int d \theta dz \dfrac{f_2 ^{1/2}f_1^{3/2 }}{z^5}\right) \nonumber
%\\
%&=V_1+V_2,
\label{f_4}
\end{align}
Similar analysis to previous cases leads to the following divergence structure for the holographic subregion complexity for this case
\begin{align}
\begin{split}
%\label{f_11}
\mathcal{C}_{c_1\times S^1}=&\dfrac{L^4 R_1 \pi}{6G_N}\Bigg[\dfrac{1}{4 \delta ^4}H^3 (1-\cos(\Omega))-\dfrac{\cos(\Omega) H}{6 \delta ^2}\\
&+\frac{1}{\delta}\Bigg(\dfrac{\cos(\Omega)}{3 h_0}+\dfrac{(-13+5 \cos (2\Omega)) \cot (\Omega) \csc(\Omega) h_0}{108 }-\dfrac{ \cos (\Omega) \cot (\Omega) f_0 h_0^2}{18}\\
&-\int _ {h_0}^{0} d h \left( \dfrac{\sin(\theta)}{\dot{h} h^3}+\dfrac{\cos (\Omega)}{3 h^2}-\dfrac{(-13+5 \cos(2 \Omega) ) \cot (\Omega) \csc (\Omega )}{108 }+\dfrac{ \cos (\Omega) \cot (\Omega) f_0 h}{9} \right)\Bigg)\\
&+\dfrac{(-13+5 \cos (2\Omega)) \cot (\Omega) \csc(\Omega) }{108 H}\log (\delta )\Bigg]+\mathrm{finite}.
\end{split}
\end{align}

%%%%%%%%%%%%%%%%%%%%%%%
\subsubsection*{Subregion $ c_1 \times S^2 $}
Next we consider the singular subregion $ c_1 \times S^2 $ in a CFT defined on $ R^4 \times S^2 $. The bulk metric is given by
\begin{align}
d s^2= \frac{L^2}{z^2}\big [dz^2 +f_1 (z) (dt^2 +d\rho ^2+ \rho^2 d \theta ^2+\rho ^2 \sin(\theta ) ^2 d \phi ^2 )+f_2 (z) R_1^2 d \Omega _2 ^2\big ],
\label{f_12}
\end{align}
where $ d\Omega _2 $ is line element over $ S^2 $ and $ f_1$ and  $f_2 $ have the following expansions 
\begin{align}
f_1 = 1+\dfrac{ z^2}{20 R_1^2}+ \dfrac{ z^4}{100 R_1^4}+\dfrac{z^6}{ R_1^6}\left(\frac{1}{1200}-\frac{1}{400}\log{z}\right)+\mathcal{O}\left(z^8\right),\nonumber\\
f_2= 1-\dfrac{z^2}{5 R_1^2}-\dfrac{7 z^4}{800 R_1^4}+\dfrac{z^6}{ R_1^6}\left(\frac{7}{4800}+\frac{1}{200}\log{z}\right)+\mathcal{O}\left(z^8\right).
\end{align}
Using the metric  (\ref{f_12}) we find the holographic complexity as
\begin{align}
\begin{split}
V(\gamma)= & 2\pi L^6 R _1^2 \Omega _2 \int d\rho d\theta dz \dfrac{f_2 f_1 ^{3/2} \rho ^2 \sin (\theta)}{z^6}\\
&= 2\pi L^6 R _1^2 \Omega _2\int d \theta dz \dfrac{f_2 f_1^{3/2}\sin (\theta)}{z^6} \int _{\rho(z,\theta)}^{H} d\rho \rho ^2 \\
&=\dfrac{ 2\pi L^6 R _1^2 \Omega _2}{3}\left(- \int d \theta dz \dfrac{f_2 f_1 ^{3/2} \rho ^3}{z^6}+ H^3 \int d \theta dz \dfrac{f_2 f_1^{3/2 }}{z^6}\right)
% \nonumber\\
%&=V_1+V_2,
\label{f_16}
\end{split}
\end{align}
Similar analysis to previous sections leads to
\begin{align}
\begin{split}
\mathcal{C}_{c_1\times S^2}^{\log }=\frac{L^5 R _1^2 \Omega _2}{12G_N}\Bigg[&\frac{3}{R_1^2}\bigg(\int _{h_0}^{0} dh\bigg[\dfrac{\sin (\theta) g_3}{\dot{h}h^2}-\dfrac{\cos (\Omega)} {80 h^2}+\dfrac{\cos (\Omega)(1+\csc ^2(\Omega) )\log (h)}{384}\\
&+\dfrac{\cos(\Omega)(-3 +27 \cot ^2(\Omega)-45 \csc ^2 (\Omega) +3840 b_3)}{15360}\bigg]+\frac{\cos (\Omega)}{80 h_0 }\\
&+\dfrac{\cos (\Omega)(1+\csc ^2 (\Omega))}{ 384}(h_0\log(h_0)-h_0)\\
&+\dfrac{\cos(\Omega)(-3 +27 \cot ^2(\Omega)-45 \csc ^2 (\Omega) +3840 b_3) h_0}{15360 }\bigg)\\
&-\frac{1}{8 R_1^2}\bigg(\int _{h_0}^{0} dh\left[\dfrac{\sin (\theta) }{\dot{h}h^3}+\dfrac{\cos (\Omega)}{4 h ^2}-\dfrac{\cos (\Omega) \csc ^2 (\Omega)(-11+5 \cos (2\Omega))}{256}\right]\\
&-\frac{\cos (\Omega)}{4 h_0}-\dfrac{\cos (\Omega)\csc ^2(\Omega) h_0 (-11+\cos(2\Omega))}{256}\bigg)\Bigg]\log(\delta).
\label{f_33}
\end{split}
\end{align}

\section{Discussions}
In this paper we studied the divergence structure of holographic subregion complexity for various singular surfaces. We showed that there are new divergences due to singularities in the subregion. More specifically we have shown that for a kink in a (2+1)-dimensional field theory and also cones $c_n$ in even dimensional field theories a new universal $\log\delta$ terms appears. In odd dimensional field theories the singularity of a cone $c_n$ gives rise to a $\log^2\delta$ divergent term. We also showed that surprisingly crease singularities of any type do not give rise to any universal term or even any new divergent term. For generalized conical singularities the situation is completely different. There are examples which new power law divergences appear but there is no new universal term due to the singularity. We found also an example, i.e. $c_1\times S^2$, with a curved locus that has a new universal term. Another type of conical singularity has $\frac{1}{\delta}\log\delta$ and $\frac{1}{\delta^2}\log\delta$ divergent terms for even and odd dual field theories respectively. The latter family is very similar to what has been recently found using `complexity=action' proposal on the Wheeler-DeWitt patch which also posses corners. We have summarized all of these results in a table in section \ref{sec:2}.  

There are several directions to follow in future works. Regarding the divergence structure of subregion complexity, the most important question is whether one can define any monotonic function from the universal terms which leads to a kind of 'c-function' in higher \textit{odd}-dimensional dual field theories?

Another interesting open question is how to generalize complexity proposals beyond Einstein gravity. Recently there have been some proposals trying to address this question (see e.g. \cite{Bueno:2016gnv}). 

A natural question about this work is how to study the role of singularities of subregions in the `complexity=action' proposal. Recently some progress have been made in \cite{Carmi:2016wjl} for spherical subregions. The authors have proposed the intersection between the ``entanglement wedge" and the corresponding WDW patch for `complexity=action' for mixed states constructed from subregions. It would be instructive to understand this proposal by considering more complicated examples.

%%%%%%%%%%%%%%%%%%%%%%%%%%%%%%%%
\section*{Acknowledgements}
We would like to thank Mohsen Alishahiha, Amin Faraji-Astaneh and Mohammad H. Vahidinia for fruitful discussions
and M. Reza Mohammadi-Mozaffar for careful reading of the manuscript.

%%%%%%%%%%%%%%%%%%%%%%%%%%%%%%%%

\end{document}